\documentclass[a4paper,11pt]{article}

\usepackage{amsmath,amsthm,amsfonts,amssymb,graphics,epsfig,color}

\usepackage[left=2cm,top=3cm,right=2cm,bottom=3cm,bindingoffset=0.5cm]{geometry}
\usepackage[latin1]{inputenc}
\usepackage[english]{babel}
\usepackage{verbatim}
\usepackage{amscd,enumerate}
\usepackage{epstopdf}
\usepackage{graphicx}
\usepackage{multicol}

\newcommand{\erf}{{\text{erf}}}
\newcommand{\Ste}{{\text{Ste}}}

\newtheorem{obs}{Remark}[section]
\newtheorem{teo}{Theorem}[section]


\makeatletter
\def\cleardoublepage{\clearpage\if@twoside \ifodd\c@page\else%
    \hbox{}%
    \thispagestyle{empty}%
    \newpage%
    \if@twocolumn\hbox{}\newpage\fi\fi\fi}
\makeatother

\def\figurename{Figure}
\makeatletter
\renewcommand{\fnum@figure}[1]{\figurename~\thefigure.}
\makeatother

\def\tablename{Table}
\makeatletter
\renewcommand{\fnum@table}[1]{\tablename~\thetable.}
\makeatother

\begin{document}

\title{Approximate solutions to the one-phase Stefan problem with non-linear temperature-dependent thermal conductivity}

\author{
Julieta Bollati$^{1}$,  Mar\'ia F. Natale$^{2}$, Jos\'e A. Semitiel$^{2}$, Domingo A. Tarzia $^{1}$\\ \\
\small{$^1$ Consejo Nacional de Investigaciones Cient\'ificas y Tecnol\'ogicas (CONICET)}\\
\small{$^2$ Depto. Matem\'atica - CONICET, FCE, Univ. Austral, Paraguay 1950} \\  
\small{S2000FZF Rosario, Argentina.} \\
\small{Email: JBollati@austral.edu.ar; JSemitiel@austral.edu.ar; DTarzia@austral.edu.ar.}
}

\maketitle

\begin{abstract}

In this chapter we consider different approximations for the one-dimensional one-phase Stefan problem corresponding to  the fusion process of a semi-infinite material with a temperature boundary condition at the fixed face and  non-linear temperature-dependent thermal conductivity. The knowledge of the exact solution of this problem, allows to compare it directly   with the approximate solutions obtained by applying the heat balance integral method, an alternative form to it and the refined balance integral method, assuming a quadratic temperature profile in space. In all cases, the analysis is carried out in a dimensionless way by the Stefan number (Ste) parameter.


\end{abstract}
\vspace{.08in}
\noindent \textbf{Keywords:}  Stefan problem, heat balance integral  method, refined integral method, temperature-dependent thermal conductivity, exact solutions

\section{Introduction}

Stefan problems model heat transfer processes that involve a change of phase. They constitute a broad field of study since they arise in a great number of mathematical and industrial significance problems. Phase-change problems appear frequently in industrial processes and other problems of technological interest \cite{AlSo}-\cite{Lu}. Reviews on the subject were given in \cite{Ta1,Ta2}. 

The heat balance integral method introduced in \cite{goodman58} is a well-known method of approximation of solutions of Stefan problems. It transforms the heat equation into an ordinary differential equation over time assuming a quadratic temperature profile in space. For these temperature profiles, different variants of this method were established by \cite{wood01}. Moreover in \cite{Hristov09}-\cite{Mosally02} the heat balance integral method is used for different temperature profiles.

In this chapter, we obtain approximate solutions to  a phase-change Stefan problem (\ref{ecq})-(\ref{s0q}) for a non-linear heat conduction equation corresponding to a semi-infinite
region $x>0$ with a thermal conductivity $k(\theta )$ given by 
\begin{equation}
k(\theta )=\frac{\rho c}{\left( a+b\theta \right) ^{2}}  \label{kq}
\end{equation}
and phase change temperature $\theta _{f}=0$. This kind of thermal
conductivity or diffusion coefficient was considered in \cite
{BaSa}-\cite{TrBr}.

The exact solution of this problem was given in \cite{NaTa}, where the temperature is the unique solution of an integral equation and the coefficient that characterizes the free boundary 
is the unique  solution of a transcendental equation. From this fact, the most remarkable aspect of this chapter lies in the comparison of the exact solution, which is difficult and cumbersome to operate, with different approaches obtained  through: the heat balance integral method, an alternative form to it  \cite{wood01}  and the refined integral method  \cite{Sadoun}.

The methods mentioned above have been developed for the non-linear diffusion equation to the  case of a linearly temperature-dependent thermal diffusivity \cite{FaHr2016} and a power-law dependent diffusivity with integer positive exponent \cite{Hr2015}, \cite{Hr2016}; obtaining closed forms of approximate solutions.

The goal of this chapter is to provide approximate solutions in order to facilitate the search of the solution and show that it is worth using approximate methods due to the small error with respect to the exact solution.  In all the applied methods the dimensionless parameter called Stefan number is defined. We take Stefan number up to 1 due to the fact that it covers most of phase change materials \cite{Solomon}.

\section{Mathematical formulation and exact solution}
\label{sec:1}
We consider a one-dimensional one-phase Stefan problem for the fusion of a semi-infinite material $x>0$ with  non-linear temperature-dependent thermal \mbox{conductivity}. This problem can be formulated mathematically in the following way:

Problem (P). Find the temperature $\theta=\theta(x,t)$ at the liquid region $0<x<s(t)$ and the evolution of the moving  separation phase given by $x=s(t)$ satisfying the following conditions 
\begin{align}
\rho c\dfrac{\partial \theta }{\partial t}&=\dfrac{\partial }{\partial x}%
\left( k(\theta )\dfrac{\partial \theta }{\partial x}\right), &0<x<s(t),\quad t>0  \label{ecq} \\
\theta &=\theta_{0}>0\;,\;& \text{ on }~ x=0,\quad t>0   \label{cf} \\
k\left( \theta  \right) \dfrac{\partial \theta }{%
\partial x} &=-\rho \lambda\dot{s}(t)\;,& \text{ on }~ x=s(t),\quad t>0 \label{csq} \\
\theta  &=0\;,\;& \text{ on }~ x=s(t),\quad t>0  \label{tflq} \\
s(0)&=0  \label{s0q}
\end{align}
where the parameters $c,\rho$ and $\lambda$ are the specific heat, the density and the latent heat of fusion of the medium respectively, all of them assumed to be positive constants. The thermal conductivity $k$ is given by (\ref{kq}), with   positive parameters $a$ and $b$.

In \cite{NaTa} was proved the existence and uniqueness of an exact solution
of the similarity type of the free boundary problem $\left( \ref{ecq}\right)
-\left( \ref{s0q}\right) $ for $t\geq t_{0}>0$ with $t_{0}$ an arbitrary
positive time when data satisfy condition $ ac=b\lambda$.

If we define the non-dimensional Stefan number by
\begin{equation}
\Ste=\frac{c\theta_{0}}{\lambda},
\end{equation}
 then we have $\Ste=\frac{b\theta_{0}}{a}$.

Now, we can write the exact
solution as \cite{NaTa}:
\begin{align}
\theta (x,t)&=\dfrac{1}{b}\left[ \dfrac{1}{\Theta (x,t)}-a\right]\;,\;0<x<s(t)\;,\;t>0, \label{theta} \\
s(t)&=\dfrac{2}{a}\xi \sqrt{t}\;,\;t>0,\label{ese}
\end{align} 
where $\Theta $ is the unique solution in variable $x$ of the following
integral equation 
\begin{equation}
\Theta (x,t)= \dfrac{1}{a}\left(1+\tfrac{\Ste}{ (1+\Ste) \erf(\Lambda)} \erf
\left( \tfrac{\int_{0}^{x}\tfrac{d\eta }{\Theta (\eta ,t)}}{2\sqrt{t}}
-\Lambda\right) \right) \;,\;0\leq x\leq s(t),
\label{soligual}
\end{equation}
for $t\geq t_{0}>0$ with $t_{0}$ an arbitrary positive time and $\xi$ is given by
\begin{equation}
\xi=\dfrac{2\Lambda \exp\left(\Lambda^2\right)}{1+\Ste}, \label{Xi}
\end{equation}
where $\Lambda$ is the unique positive solution to the following equation 
\begin{equation}
z\exp (z^{2})\erf(z)=\frac{\Ste}{%
\sqrt{\pi }}\;\;\;,\;\;\;z>0.
\label{experf}
\end{equation}
 
\begin{obs} In \cite{NaTa} was proved that  the integral equation (\ref{soligual}) is equivalent to solve the following Cauchy differential problem in variable $x$:
\begin{equation}
\left\{
     \begin{array}{ll}
        \dfrac{\partial Y}{\partial x}(x,t)=\dfrac{a}{2\sqrt{t}\left[1+\frac{\Ste}{(1+\Ste)\erf\left(\Lambda\right)}\erf\left(Y(x,t)\right)\right]}, \quad 0<x<s(t), \ t>0,\\     
       Y(0,t)=-\Lambda, \\               						\end{array}
                        \right.
\end{equation}  where 
\begin{equation}
Y(x,t)=\dfrac{\int_{0}^{x}\tfrac{d\eta}{\Theta (\eta ,t)}}{2\sqrt{t}}
-\Lambda
\end{equation}
with a positive parameter $t\geq t_{0}>0$ and $\Lambda$ the unique solution of (\ref{experf}).
\end{obs}

\section{Heat balance integral methods}
\label{sec:3}
As one of the mechanisms for the heat conduction is the diffusion, the excitation at the fixed face $x=0$ (for example, a temperature, a flux or a convective condition) does not spread instantaneously to the material $x>0$. However, the effect of the fixed boundary condition can be perceived  in a bounded interval $\left[0,\delta(t)\right]$ (for every time $t>0$) outside of which the temperature remains equal to the initial temperature. The heat balance integral method presented in \cite{goodman58} 
established the existence of a function $\delta=\delta(t)$ that measures the depth of the thermal layer. In problems with a phase of change, this layer is assumed as the free boundary, i.e $\delta(t)=s(t)$.

From equation (\ref{ecq}) and conditions (\ref{csq}) and (\ref{tflq}) we obtain the new condition:
\begin{equation} \label{CondStefanAprox}
k\left(\theta \right)\left(\dfrac{\partial \theta}{\partial x} \right)^2~=\dfrac{\lambda}{c}\dfrac{\partial}{{\partial x}} \left( k(\theta)\frac{\partial \theta}{\partial x} \right) \qquad \text{ on }\quad x=s(t),\ \ t>0.
\end{equation}

From equation (\ref{ecq}) and conditions (\ref{cf}), (\ref{csq}) and (\ref{tflq}) we obtain the integral condition:
\begin{eqnarray}
\dfrac{d}{dt} \int\limits_{0}^{s(t)} \theta(x,t) dx&=&\int\limits_{0}^{s(t)} \dfrac{\partial\theta }{\partial t}(x,t)dx +\theta(s(t),t)\dot{s}(t) \nonumber\\ 
&=&\dfrac{1}{\rho c}\int\limits_0^{s(t)} \dfrac{\partial}{\partial x} \left(k\left(\theta(x,t)\right)\dfrac{\partial\theta}{\partial x}(x,t) \right)dx \nonumber \\
&=&\dfrac{-1}{\rho c}\left[\rho \lambda \dot{s}(t)+k\left(\theta_0\right)\dfrac{\partial\theta}{\partial x}(0,t) \right].\label{EcCalorAprox}
\end{eqnarray}

The classical heat balance integral method introduced in \cite{goodman58} proposes to approximate problem (P) through the resolution of a problem that arises on replacing the equation (\ref{ecq}) by the equation                     (\ref{EcCalorAprox}), and the condition (\ref{csq}) by the condition (\ref{CondStefanAprox}); that is, the resolution of the approximate problem defined as follows: conditions (\ref{cf}), (\ref{tflq}), (\ref{s0q}), (\ref{CondStefanAprox}) and (\ref{EcCalorAprox}).

In \cite{wood01}, a variant of the classical heat balance integral method was proposed by replacing equation (\ref{ecq}) by condition (\ref{EcCalorAprox}), keeping all others conditions of the problem (P) equals;  that is, the resolution of an approximate problem defined as follows: conditions (\ref{cf}),(\ref{csq}),(\ref{tflq}),(\ref{s0q}) and (\ref{EcCalorAprox}).

From equation (\ref{ecq}) and conditions (\ref{cf}) and (\ref{tflq}) we also obtain:
\begin{eqnarray}
\int\limits_0^{s(t)} \int\limits_0^x \dfrac{\partial\theta}{\partial t}(\eta,t) d\eta dx &=& \int\limits_0^{s(t)}  \int\limits_0^x \dfrac{1}{\rho c} \dfrac{\partial}{\partial \eta} \left(k\left(\theta(\eta,t)\right)\dfrac{\partial\theta}{\partial \eta}(\eta,t) \right)d\eta dx \nonumber \\
&=& \int\limits_0^{s(t)} \dfrac{1}{\rho c} \left[k\left(\theta(x,t)\right)\dfrac{\partial\theta}{\partial x}(x,t)- k\left(\theta_0\right)\dfrac{\partial\theta}{\partial x}(0,t) \right] dx \nonumber \\
&=& \dfrac{1}{\rho c} \int\limits_0^{s(t)} \rho c \dfrac{\dfrac{\partial\theta}{\partial x}(x,t)}{(a+b\theta(x,t))^{2}} dx-\dfrac{k\left(\theta_0\right)}{\rho c} \dfrac{\partial\theta}{\partial x}(0,t)s(t)
\nonumber \\
&=&-\dfrac{\theta_0\left(1+\Ste\right)+\dfrac{\partial \theta}{\partial x}(0,t)s(t)}{a^2\left(1+\Ste\right)^2}. \label{EcCalorAproxRIM}
\end{eqnarray}

The refined  integral method introduced in \cite{Sadoun} proposes to approximate problem (P) through the resolution of the approximate problem that arises by replacing equation (\ref{ecq}) by (\ref{EcCalorAproxRIM}), keeping all others conditions of the problem (P) equals. It is defined as follows: conditions (\ref{cf}), (\ref{csq}), (\ref{tflq}), (\ref{s0q}) and (\ref{EcCalorAproxRIM}).

For solving the approximate problems previously defined  we propose a quadratic temperature profile in space as follows:
\begin{equation}\label{Perfil}
\widetilde{\theta}(x,t)=\widetilde{A}\theta_{0}\left( 1-\dfrac{x}{\widetilde{s}(t)}\right)+\widetilde{B}\theta_{0}\left(  1-\dfrac{x}{\widetilde{s}(t)}\right)^2, 
\end{equation}
where $\widetilde{\theta}$ and $\widetilde{s}$ are approximations of $\theta$ and $s$ respectively.

Taking advantage of the fact of having the exact temperature  of the problem (P), it is physically reasonable to impose that the approximate temperature given by (\ref{Perfil}) behaves in a similar manner than the exact one given by (\ref{theta}); that is: its sign, monotony and convexity in space. As $\theta$ verifies the following properties: 
\begin{align}
& \theta(x,t)>0,\\
& \dfrac{\partial \theta}{\partial x}(x,t) = -\dfrac{\theta_0}{a\Ste}\frac{1}{\Theta^2(x,t)}\dfrac{\partial \Theta}{\partial x}(x,t)<0, \\
& \dfrac{\partial^2 \theta}{\partial x^2}(x,t) = -\tfrac{\theta_0}{a\Ste}\left(-\tfrac{2}{\Theta^3(x,t)}\tfrac{\partial \Theta}{\partial x}(x,t)+\tfrac{1}{\Theta^2(x,t)}\tfrac{\partial^2 \Theta}{\partial x^2}(x,t)\right)>0,
\end{align}
on $0<x<s(t)$, $t>0$, we enforce the following conditions on $\widetilde{\theta}$:
\begin{align}
& \widetilde{\theta}(x,t)> 0,\\
&\dfrac{\partial \widetilde{\theta}}{\partial x}(x,t) = -\dfrac{\theta_{0}}{\widetilde{s}(t)} \left(\widetilde{A}+2\widetilde{B}\left(1 -\dfrac{x}{\widetilde{s}(t)}\right) \right)<0, \\
&\dfrac{\partial ^2\widetilde{\theta}}{\partial x^2}(x,t) = \dfrac{2\widetilde{B}\theta_{0}}{\widetilde{s}^2(t)}>0, \qquad
\end{align}
for all $0<x<\widetilde{s}(t)$, $t>0$. Therefore, we obtain that both constants $\widetilde{A}$ and $\widetilde{B}$ must be positive.

\subsection{Approximate solution using the classical heat balance integral method}

The classical heat balance integral method proposes to approximate problem (P) through the resolution of the approximate problem defined in Sect. \ref{sec:3}, that is:

Problem (P1). Find the temperature $\theta_{1}=\theta_{1}(x,t)$ at the liquid region $0<x<s_{1}(t)$ and the location of the free boundary $x=s_{1}(t)$ such that:
\begin{align} 
\tfrac{d}{dt} \int\limits_{0}^{s_{1}(t)} \theta_{1}(x,t) dx &=\tfrac{-1}{\rho c}\left[\rho \lambda \dot{s}_{1}(t)+k\left(\theta_{0}\right)\tfrac{\partial\theta_{1}}{\partial x}(0,t) \right], & 0<x<s_{1}(t), \label{EcCalorP1}\\
\theta_{1}&=\theta_{0},& \text{ on } ~ x=0,\ \ \ \ \ \label{CondConvP1}\\
k\left(\theta_1 \right)\left(\dfrac{\partial \theta_1}{\partial x} \right)^2&=\dfrac{\lambda}{c}\dfrac{\partial}{{\partial x}} \left( k(\theta_1)\frac{\partial \theta_1}{\partial x} \right),&\text{ on } ~ x=s_1(t), \label{PseudoStefan}\\
\theta_{1}&= 0, & \text{ on } ~ x=s_1(t),\label{TempFronteraP1}\\
s_{1}(0)&=0. \label{FrontinicialP1}
\end{align} 

By proposing the following quadratic temperature profile in space:
\begin{equation}
\theta_{1}(x,t)= \theta_{0}A_{1}\left(1-\frac{x}{s_{1}(t)}\right) +\theta_{0}B_{1}\left(1-\dfrac{x}{s_{1}(t)}\right)^{2}, \quad 0<x<s_1(t), \ ~ t>0,\label{TempP1}
\end{equation}
 the free boundary is obtained of the form:
\begin{equation}
s_{1}(t)=\dfrac{2}{a} \xi_{1} \sqrt{t},\quad t>0,\label{FrontP1}
\end{equation}
where the constants $A_{1}, B_{1}$ y $\xi_{1}$ will be determined from the conditions (\ref{EcCalorP1}), (\ref{CondConvP1}) and (\ref{PseudoStefan}) of the problem (P1).
Because of (\ref{TempP1}) and (\ref{FrontP1}), the conditions (\ref{TempFronteraP1}) and (\ref{FrontinicialP1}) are immediately satisfied. From conditions (\ref{EcCalorP1}) and (\ref{CondConvP1}) we obtain:
\begin{equation}\label{A1}
A_{1}=\dfrac{2\left[3\Ste-(1+\Ste)^2\xi_1^2(\Ste+3) \right]}{\Ste\left[(1+\Ste)^2\xi_1^2+3\right]},
\end{equation}

\begin{equation}\label{B1}
B_{1}=\dfrac{3\left[-\Ste+(1+\Ste)^2\xi_1^2(\Ste+2) \right]}{\Ste\left[(1+\Ste)^2\xi_1^2+3\right]}.
\end{equation}

From the fact that $A_1>0$ and $B_1>0$ we obtain that $0<\xi_1<\xi^{\max}$ and $\xi_1>\xi^{\min}>0$, respectively where:
\begin{equation}
\xi^{\min}=\sqrt{\dfrac{\Ste}{(1+\Ste)^2(2+\Ste)}}~,  \qquad \xi^{\text{max}}=\sqrt{\dfrac{3\Ste}{(1+\Ste)^2(3+\Ste)}}~.
\end{equation}

Since $A_{1}$ and $B_{1}$ are defined from the parameters $\xi_{1}$ and $\Ste$, condition (\ref{PseudoStefan}) will be used to find the value of $\xi_{1}$. In this way, it turns out that $\xi_{1}$ must be a positive solution of the fourth degree polynomial equation:
\begin{eqnarray} 
Q_{1}(z)&&:= \left(1+\Ste\right)^{4}\left(2\Ste^{2}+11\Ste+16\right)z^{4}\nonumber\\
&&-2\left(1+\Ste\right)^{2}\left(6\Ste^{2}+19\Ste+3\right)z^{2} \nonumber\\
&&+3\Ste\left(1+6\text{Ste} \right)=0,\qquad \xi^{\min}<z<\xi^{\max}~.
\end{eqnarray}
It is easy to see that $Q_{1}$ has only two positive roots. In addition:
\begin{align}
&Q_1(\xi^{\min})=\tfrac{2\Ste^2\left(2\Ste+3 \right)^2}{\left(2+\Ste \right)^2}>0,\\
&Q_1(\xi^{\max})=-\tfrac{3\Ste\left(2\Ste+3 \right)^2}{\left(3+\Ste \right)^2}<0,\\
& Q_1(+\infty)=+\infty.
\end{align}
Therefore $Q_1$ has a unique root in $\left(\xi^{\min},\xi^{\max}\right)$ and it is  given explicitly by 
\begin{equation} \label{Xi-1Explicita}
\xi_1=\left(\tfrac{{\left(\mathrm{Ste} + 1\right)}^2\, \left(6\, {\mathrm{Ste}}^2 + 19\, \mathrm{Ste} + 3\right) - \sqrt{6\, \mathrm{Ste} + 1}\, \left(2\, {\mathrm{Ste}}^2 + 5\, \mathrm{Ste} + 3\right)}{{\left(\mathrm{Ste} + 1\right)}^4\, \left(2\, {\mathrm{Ste}}^2 + 11\, \mathrm{Ste} + 16\right)}\right)^{1/2}.
\end{equation}

All the above analysis can be summarized in the following result:
\begin{teo}
The solution to the problem (P1), for a quadratic profile in space, is given by (\ref{TempP1}) and (\ref{FrontP1}) where the positive constants $A_1$ and $B_1$ are defined by (\ref{A1}) and (\ref{B1}) respectively and $\xi_1$ is given explicitly by (\ref{Xi-1Explicita}).
\end{teo}

As this approximate method is designed as a technique for tracking the location of the free boundary, the comparisons between the approximate solutions and the exact one are made on the free boundary thought the coefficients that characterizes them (Fig.1). Generally for most phase-change materials candidates over a realistic temperature, the Stefan number will not exceed 1 \cite{Solomon}. From this, in order to analyse the accuracy of the approximate solution we compare the dimensionless coefficients  $\xi_1$ with the exact coefficient $\xi$ given by (\ref{Xi}) for Ste $<1$. Moreover, in Fig.2, we show the temperature profile of the approximate solution and the exact one at $t=10s,$  for the parameters  $\Ste=0.4, a=1\ \sqrt{s}/m $ and $\theta_{0}=3^{\circ}C$.

\begin{figure*}[h]
\begin{center}
\includegraphics[width=3.5in]{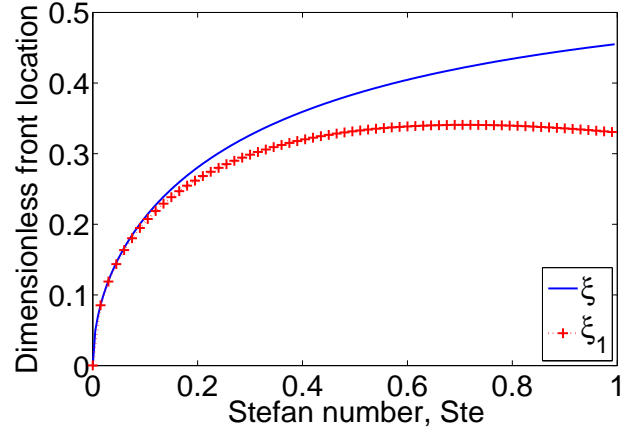} 
\end{center}
\caption{Plot of $\xi$ and $\xi_1$ against $\Ste$.}
\label{Fig1}
\end{figure*}

\begin{center}

\begin{figure*}[h]

\begin{center}
 \includegraphics[width=3.5in]{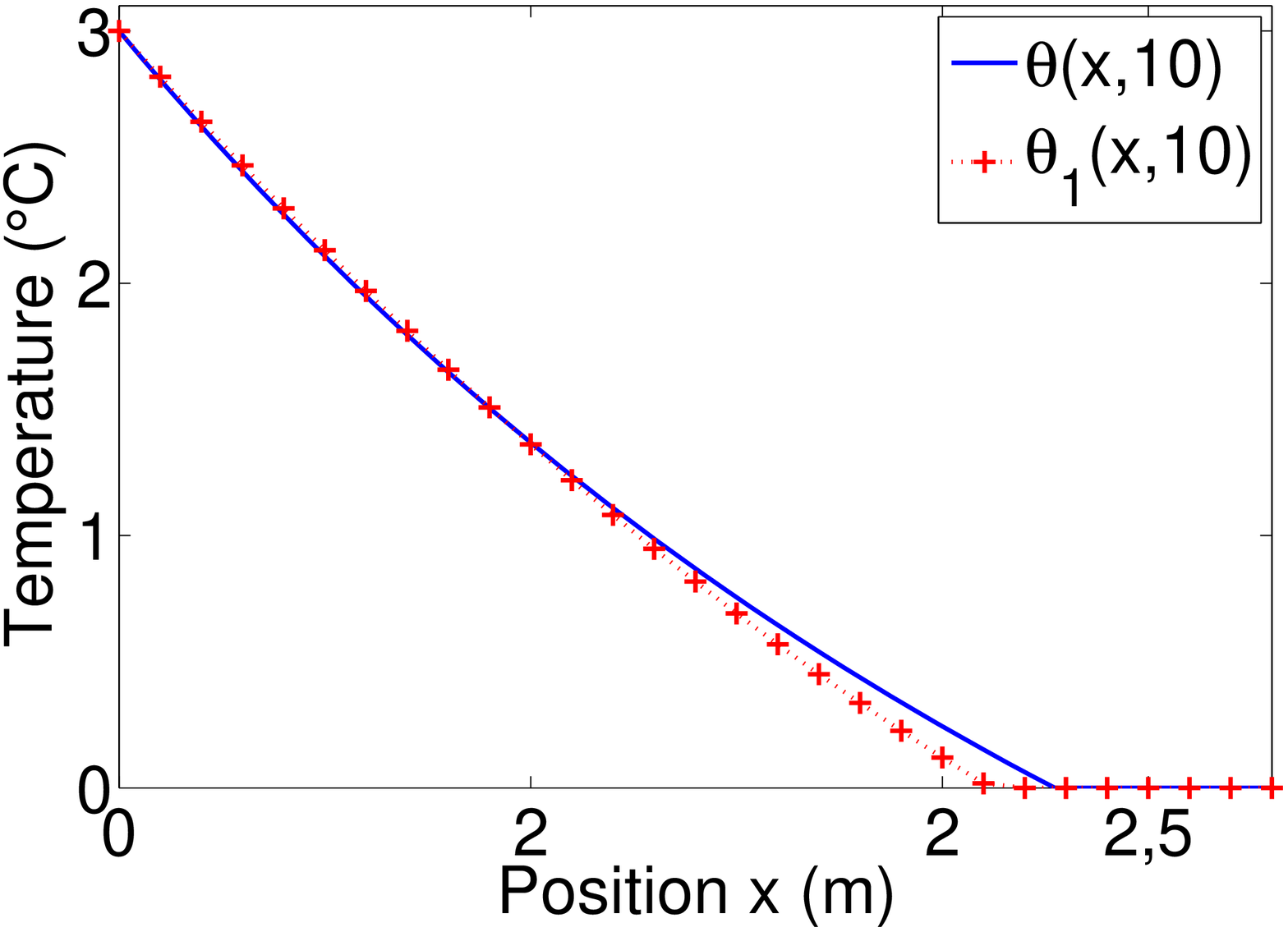} 
\end{center}
\caption{Plot of $\theta$ and $\theta_1$ against $x$ at $t=10$ s for $\Ste=0.4$, $a=1$ $\sqrt{s}/m$, $\theta_{0}=3^{\circ}C$. }
\label{Fig2}
\end{figure*}
\end{center}

\subsection{Approximate solution using an alternative of the heat balance integral method}
An alternative method of the classical heat balance integral method proposes to approximate problem (P) through the resolution of the approximate problem defined in Sect. \ref{sec:3}, that is:

\medskip

Problem (P2). Find the temperature $\theta_{2}=\theta_{2}(x,t)$ at the liquid region $0<x<s_{2}(t)$ and  the location of the free boundary $x=s_{2}(t)$ such that:
\begin{align}
\tfrac{d}{dt} \int\limits_{0}^{s_{2}(t)} \theta_{2}(x,t) dx&=\frac{-1}{\rho c}\left[\rho \lambda \dot{s}_{2}(t)+k\left(\theta_{0}\right)\tfrac{\partial\theta_{2}}{\partial x}(0,t) \right],  & 0<x<s_{2}(t), \label{EcCalorP2}\\
\theta_{2}&=\theta_{0}>0, &\text{ on } ~ x=0,\ \ \ \ \  \label{CondConvP2}\\
k\left(\theta_{2}\right)\dfrac{\partial\theta_{2}}{\partial x}&=-\rho \lambda \dot{s}_{2}(t),  & \text{ on } ~ x=s_2(t), \label{CondStefanP2} \\
\theta_{2}&= 0, & \text{ on } ~ x=s_2(t), \label{TempFronteraP2}\\
s_{2}(0)&=0.& \label{FrontinicialP2}
\end{align} 

The solution of the problem (P2), for a quadratic temperature profile in space, is obtained by
\begin{align}
\theta_{2}(x,t)&= \theta_{0}A_{2}\left(1-\dfrac{x}{s_{2}(t)}\right) +\theta_{0}B_{2}\left(1-\dfrac{x}{s_{2}(t)}\right)^{2},\quad 0<x<s_2(t),~t>0, \label{TempP2}\\
s_{2}(t)&=\dfrac{2}{a} \xi_{2} \sqrt{t},\quad t>0, \label{FrontP2}
\end{align}
where the constants $A_{2}, B_{2}$ y $\xi_{2}$ will be determined from the conditions (\ref{EcCalorP2}), (\ref{CondConvP2}) and (\ref{CondStefanP2}) of the problem (P2). The conditions (\ref{TempFronteraP2}) and (\ref{FrontinicialP2}) are immediately satisfied. From conditions (\ref{CondConvP2}) and (\ref{CondStefanP2}), we obtain:
\begin{equation}\label{A2}
A_{2}=\dfrac{2\xi_{2}^{2}}{\Ste},
\end{equation}
\begin{equation}\label{B2}
B_{2}=1-\dfrac{2\xi_{2}^{2}}{\Ste}.
\end{equation}

As we know, the constants $A_2$ and $B_2$ must be positive then we have $0<\xi_2<\sqrt{\frac{\Ste}{2}}$.

Moreover, as in the previous problem (P1), the constants $A_{2}$ and $B_{2}$ are expressed as a function of the parameters
$\xi_{2}$ and $\Ste$, and using condition (\ref{EcCalorP2}) the coefficient $\xi_{2}$ must be a positive solution of the fourth degree polynomial equation given by:
\begin{eqnarray}
\left(1+\Ste\right)^{2} z^{4}+\left(6+7 \Ste +5 \Ste^{2}+\Ste^{3}\right)z^{2}-3 \Ste=0, \label{Xi-2}
\end{eqnarray}
for $0<z<\sqrt{\tfrac{\Ste}{2}}$.\\
Then, it is easy to see that the above equation has a unique solution given explicitly by
\begin{equation}
\xi_2=\left(\tfrac{-\left(6+7\Ste+5\text{Ste}^2+\text{Ste}^3\right)+\sqrt{\left(6+7\text{Ste}+5\text{Ste}^2+\text{Ste}^3\right)^2+12\text{Ste}\left(1+\text{Ste}\right)^2}}{2\left(1+\text{Ste}\right)^2}\right)^{1/2}. \label{Xi-2Explicita}
\end{equation}

All the above analysis can be summarized in the following result:
\begin{teo}
The solution to the problem (P2), for a quadratic profile in space, is given by (\ref{TempP2}) and (\ref{FrontP2}) where the positive constants $A_2$ and $B_2$ are defined by (\ref{A2}) and (\ref{B2}) respectively and $\xi_2$ is given explicitly by (\ref{Xi-2Explicita}).
\end{teo}

\begin{figure*}[h!]
\begin{center}
\begin{center}
\includegraphics[width=3.6in]{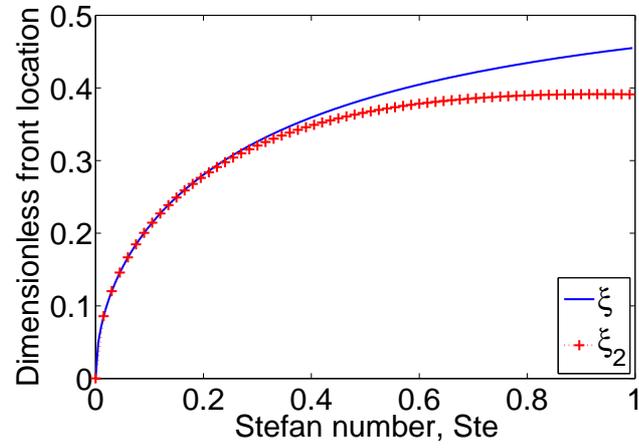} 
\end{center}
\end{center}
\caption{Plot of $\xi$ and $\xi_2$ against $\Ste$.}
 \label{Fig3}
 
\end{figure*}

\medskip

\begin{figure*}[h!]
\begin{center}
 \begin{center}
 \includegraphics[width=3.6in]{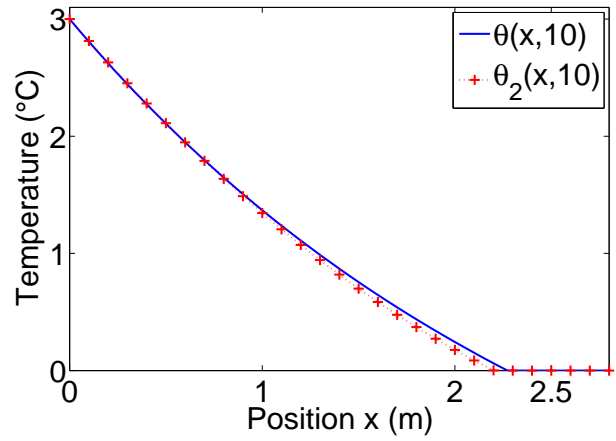} 
 \end{center}
\end{center}
\caption{Plot of $\theta$ and $\theta_2$ against $x$ at $t=10$ s for $\Ste=0.4$, $a=1$ $\sqrt{s}/m$, $\theta_{0}=3^{\circ}C$.}
 \label{Fig4}
 
\end{figure*}

Fig. 3 shows, for Stefan values up to 1, how the dimensionless coefficient $\xi_2$, which characterizes the location of the free boundary $s_2$, approaches the coefficient $\xi$, corresponding  to the exact free boundary $s$.  Moreover, in Fig.4, we show the temperature profile of the approximate solution and the exact one at $t=10s$ for the parameters $  \Ste=0.4, a=1 ~ \sqrt{s}/m $ and $\theta_{0}=3^{\circ}C$.

\subsection{Approximate solution using a refined balance heat integral method}
The refined integral method proposes to approximate problem (P) through the resolution of an approximate problem formulated in Section 3, that is:

\medskip

Problem (P3). Find the temperature $\theta_{3}=\theta_{3}(x,t)$ at the liquid region $0<x<s_{3}(t)$ and the location of the free boundary $x=s_{3}(t)$ such that:
\begin{align} 
\int\limits_0^{s_{3}(t)} \int\limits_0^x \tfrac{\partial\theta_{3}}{\partial t}(\eta,t) d\eta dx &=\tfrac{-\theta_0\left(1+\Ste\right)-\tfrac{\partial \theta_3}{\partial x}(0,t)s_3(t)}{a^2\left(1+\Ste\right)^2},  & 0<x<s_{3}(t),  \label{EcCalorP3}\\
\theta_{3}&=\theta_{0}>0, &\text{ on } ~ x=0 \ \ \ \ \ \, \label{CondConvP3}\\
k\left(\theta_{3}\right)\dfrac{\partial\theta}{\partial x}&=-\rho \lambda \dot{s}_{3}(t) , \quad & \text{ on } ~ x=s_3(t), \label{CondStefanP3} \\
\theta_{3}&= 0,& \text{ on } ~ x=s_3(t), \label{TempFronteraP3}\\
s_{3}(0)&=0. &\label{FrontinicialP3}
\end{align} 

The solution of the problem (P3) for a quadratic temperature profile in space is given by:
\begin{equation}\label{TempP3}
\theta_{3}(x,t)= \theta_{0}A_{3}\left(1-\dfrac{x}{s_{3}(t)}\right) +\theta_{0}B_{3} \left(1-\dfrac{x}{s_{3}(t)}\right)^{2}, \quad 0<x<s_3(t),~t>0
\end{equation}
and the free boundary is obtained of the form:
\begin{equation}
s_{3}(t)=\dfrac{2}{a} \xi_{3} \sqrt{t},\quad t>0,\label{FrontP3}
\end{equation}
where the constants $A_{3}$ , $B_{3}$ y $\xi_{3}$ will be determined from the conditions (\ref{EcCalorP3}), (\ref{CondConvP3}) and (\ref{CondStefanP3}) of the problem (P3).

From conditions  (\ref{CondConvP3}) and (\ref{CondStefanP3}) we obtain:
\begin{equation}
A_{3}=\dfrac{2 \xi_{3}^{2}}{\Ste},\label{A3}
\end{equation}
\begin{equation}
B_{3}=1-\frac{2}{\Ste}\xi_{3}^{2}.\label{B3}
\end{equation}

As is already know $A_3 >0$ and $B_3>0$, thus we obtain that $0< \xi_3 < \sqrt{\frac{\Ste}{2}}$. Moreover, since $A_{3}$ and $B_{3}$ are defined from the parameter $\xi_{3}$, condition (\ref{EcCalorP3}) will be used to find the value of $\xi_{3}$. In this way it turns out that $\xi_{3}$ must be a positive solution of the 
second degree polynomial equation:
 \begin{eqnarray}
\left(\Ste^3+2\Ste^2+\Ste+6\right) z^{2}+3\Ste (\Ste-1)=0, \qquad 0< z < \sqrt{\frac{\Ste}{2}}. 
\end{eqnarray}

Then, it is easy to see that the above equation has a unique solution if and only if  $\text{Ste}<1$ which is given explicitly by:
\begin{equation}
\xi_{3}=\left(\dfrac{3\text{Ste}(1-\text{Ste})}{\text{Ste}^3+2\text{Ste}^2+\text{Ste}+6}\right)^{1/2}.\label{Xi-3Explicita}
\end{equation}

All the above analysis can be summarized in the following result:
\begin{teo}
The solution to the problem (P3), for a quadratic profile in space, is given by (\ref{TempP3}) and (\ref{FrontP3}) where the positive constants $A_3$ and $B_3$ are defined by (\ref{A3}) and (\ref{B3}) respectively and $\xi_3$ is given explicitly by (\ref{Xi-3Explicita}).
\end{teo}

Therefore for every Ste $<1$, we plot the numerical value of the dimensionless coefficient $\xi_3$ against the exact coefficient $\xi$ (Fig.5). It can be seen that the refined integral method results  in good agreement with the exact solution of the problem (P), only for lower values of Stefan number.  Moreover, in Fig.6, we show the temperature profile of the approximate solution and the exact one at $t=10s$ for the parameters $ \Ste=0.4, a=1 ~ \sqrt{s}/m$ and $\theta_{0}=3^{\circ}C$.

\begin{figure*}[h!]
\begin{center}
\includegraphics[width=3.5in]{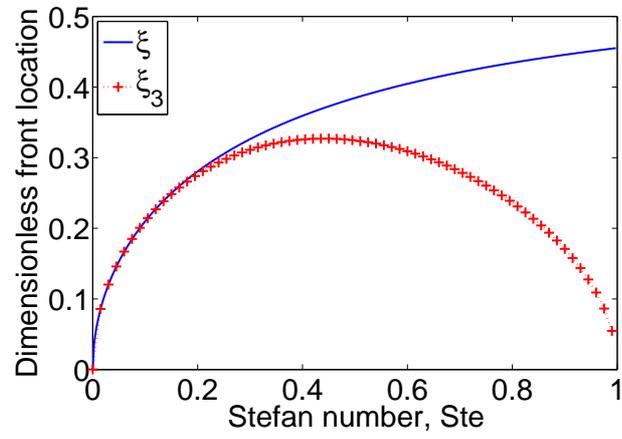} 
\end{center}
\caption{{\small Plot of $\xi$ and $\xi_3$ against $\Ste$.}}
\label{Fig5}
\end{figure*}

\begin{figure*}[h!]
\begin{center}
 \includegraphics[width=3.5in]{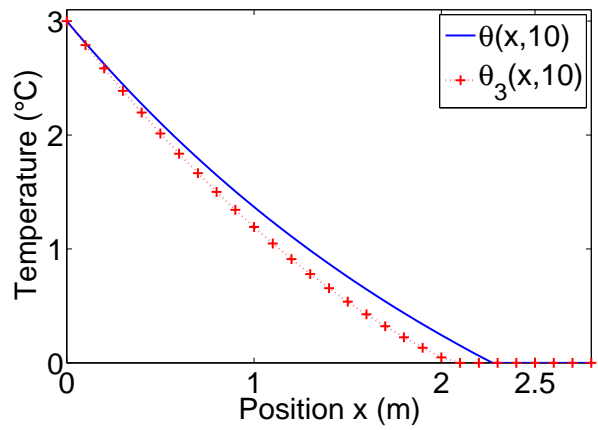} 
\end{center}
\caption{{\small Plot of $\theta$ and $\theta_3$ against $x$ at $t=10$ s for $\Ste=0.4$, $a=1$ $\sqrt{s}/m$, $\theta_{0}=3^{\circ}C$.}}
\label{Fig6}
\end{figure*}

\newpage

\section{Comparisons between solutions}

In the previous sections we have applied 3 different approximate methods (heat balance integral method (HBIM), an alternative of the HBIM and the refined integral method (RIM)) for solving a Stefan problem with a  non-linear temperature-dependent thermal conductivity.

For each of this methods, i.e. for each Problem (Pi), $i=1,2,3$ it has been plotted the dimensionless coefficient that characterizes the approximate free front $\xi_i$ versus the coefficient $\xi$ corresponding to the exact moving boundary  of problem (P). (Fig.1,3,5)

The aim of this section is to present, for different  Stefan numbers, the numerical value of the exact coefficient $\xi$ given by (\ref{Xi}) and the approximate coefficients $\xi_1$, $\xi_2$ and $\xi_3$ given by the analytical expressions (\ref{Xi-1Explicita}), (\ref{Xi-2Explicita}) and (\ref{Xi-3Explicita}) \mbox{respectively}. Those calculations will allow us not only to compare the approximate solutions with the exact one but also to compare the different approaches between them in order to show which technique gives the best agreement. With that purpose we display in Table 1, for different values of Ste, the exact dimensionless free front $\xi$, the approximate dimensionless free front $\xi_i$ and the porcentual relative error $E_{\text{rel}}(\xi_i)=100\left\vert\frac{\xi-\xi_i}{\xi} \right\vert$, $i=1,2,3$.

It may be noticed in Table 1 that the relative error committed in each approximate technique increases when the Stefan number becomes greater reaching the percentages 21\%, 14\% and 100\% for the problems (P1), (P2) and (P3) \mbox{respectively}.
From this fact, we study the behaviour of the different approaches for $\Ste<<1$ (Table 2). In this case the relative errors for problem (P1), (P2) and (P3) does not exceed 0.5\%.

\begin{table}[h!!]
\caption{Dimensionless free front coefficients and its relative errors.}
\label{tab:1}  
\begin{center}
\begin{tabular}{cc|cc|cc|cc}
\hline
Ste      & $\xi$     &   $\xi_1$ & $E_{\text{rel}}(\xi_1)$       &   $ \xi_2 $  & $E_{\text{rel}}(\xi_2)$      &   $\xi_3$ & $E_{\text{rel}}(\xi_3)$   \\
\hline

  	0.1  &  0.2099  &  0.2099 &  0.037 \% &  0.2100 &  0.018 \%  &   0.2100  & 0.042 \% \\
    0.2  &  0.2805  &  0.2754 &  1.803 \% &  0.2788 &  0.608 \%  &   0.2763  & 1.498 \% \\
    0.3  &  0.3262  &  0.3126 &  4.194 \% &  0.3207 &  1.697 \%  &   0.3112  & 4.622 \% \\
    0.4  &  0.3593  &  0.3348 &  6.809 \% &  0.3481 &  3.110 \%  &   0.3258  & 9.330 \% \\
    0.5  &  0.3846  &  0.3482 &  9.470 \% &  0.3663 &  4.741 \%  &   0.3244  & 15.63 \% \\
    0.6  &  0.4046  &  0.3557 &  12.09 \% &  0.3782 &  6.515 \%  &   0.3091  & 23.60 \% \\
    0.7  &  0.4209  &  0.3593 &  14.63 \% &  0.3856 &  8.375 \%  &   0.2802  & 33.41 \% \\
    0.8  &  0.4343  &  0.3602 &  17.07 \% &  0.3897 &  10.28 \%  &   0.2364  & 45.58 \% \\
    0.9  &  0.4457  &  0.3592 &  19.41 \% &  0.3913 &  12.20 \%  &   0.1709  & 61.66 \% \\
    1.0  &  0.4554  &  0.3568 &  21.63 \% &  0.3911 &  14.11 \%  &        0  & 100.0 \% \\

    \hline
\end{tabular}
\end{center}
\end{table}

\begin{table}[h!!!]
\caption{Dimensionless free front coefficients and its relative errors.}
\label{tab:2}  
\begin{center}
\begin{tabular}{cc|cc|cc|cc}
\hline
Ste        & $\xi$    &   $\xi_1$    & $E_{\text{rel}}(\xi_1)$       &  $\xi_2$   & $E_{\text{rel}}(\xi_2)$      &  $\xi_3$ & $E_{\text{rel}}(\xi_3)$   \\
\hline
   0.01 &    0.0702  &  0.0703  &  0.142 \% &   0.0703  &  0.037 \% &    0.0703 &   0.075 \% \\
    0.02 &    0.0987  &  0.0989  &  0.241 \% &   0.0988  &  0.066 \% &    0.0988 &   0.135 \% \\
    0.03 &    0.1201  &  0.1205  &  0.302 \% &   0.1202  &  0.086 \% &    0.1203 &   0.178 \% \\
    0.04 &    0.1378  &  0.1382  &  0.329 \% &   0.1379  &  0.099 \% &    0.1381 &   0.206 \% \\
    0.05 &    0.1531  &  0.1536  &  0.326 \% &   0.1532  &  0.103 \% &    0.1534 &   0.219 \% \\
    0.06 &    0.1666  &  0.1671  &  0.296 \% &   0.1668  &  0.101 \% &    0.1670 &   0.215 \% \\
    0.07 &    0.1789  &  0.1793  &  0.242 \% &   0.1790  &  0.090 \% &    0.1792 &   0.196 \% \\
    0.08 &    0.1901  &  0.1904  &  0.167 \% &   0.1902  &  0.073 \% &    0.1904 &   0.160 \% \\
    0.09 &    0.2004  &  0.2005  &  0.073 \% &   0.2005  &  0.049 \% &    0.2006 &   0.109 \% \\
    \hline
\end{tabular}
\end{center}
\end{table}

\bigskip

\begin{figure*}[h!]

\begin{center}
\includegraphics[scale=0.40]{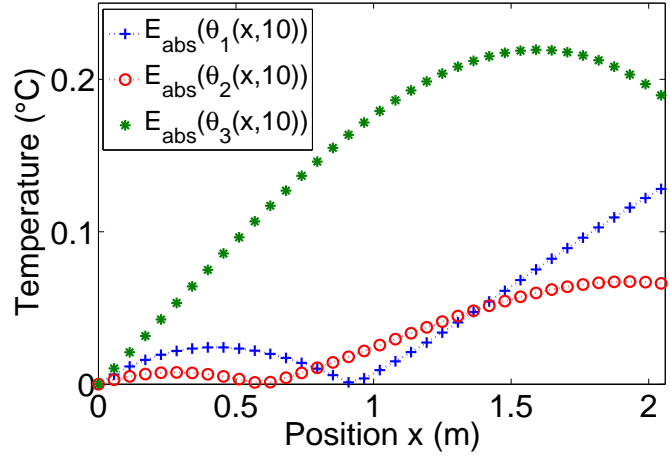}
\end{center}

\caption{{\small Temperatures absolute errors against $x$ at  $t=10s$ for $ \Ste=0.4, a=1 \sqrt{s}/m$ and $\theta_{0}=3^{\circ}C$.}}
\label{Fig8}

\end{figure*}

In Fig. \ref{Fig8} we present a comparison of the absolute errors of the  approximate temperatures given by $E_{\text{abs}}(\theta_i(x,t))=\left\vert \theta(x,t)-\theta_i(x,t)\right\vert$ , $i=1,2,3$ against the position $x$, at $t=10s,  \Ste=0.4, a=1 \sqrt{s}/m$ and $\theta_{0}=3^{\circ}C$.

\newpage

\begin{obs}
In order to compare the absolute errors of the different approaches in a common domain, in Fig. \ref{Fig8},   we plot up to $x=s_3(10)=\min\left\lbrace s_1(10),s_2(10),s_3(10)\right\rbrace$.
\end{obs}

\section{Conclusion}

In this chapter it was provided an overview of the  popular approaches such as  HBIM, RIM,  for the  case of a one-dimensional one-phase Stefan problem (P) with  a  non-linear temperature-dependent thermal conductivity as the novel feature.

It must be emphasized that the fact of having the exact solution of problem (P) has allowed  us to measure the accuracy of the approximate techniques applied throughout this chapter. Comparisons with known solution have been made in all cases and all solutions have been presented in graphical form.

It has been observed that as the Stefan number increases, the  coefficients that characterizes the free approximate boundaries move away from the exact one. However, for $\Ste<<1$, the three approaches commit a relative error that does not exceed 0.5\%.

In all the analysed cases, it could be concluded that the alternative technique of HBIM given by problem (P2) is significantly more accurate than the others.

\section*{Acknowledgements}

The present work has been partially sponsored by Projects PIP No. 112-20150100275CO CONICET-Univ. Austral and PICTO Austral 2016 No. 0090, Rosario (Argentina).

\bibliographystyle{amsplain}

\label{lastpage-01}

%

\end{document}